\documentstyle[epsf,12pt]{article}

\begin{document}
\thispagestyle{empty}
\begin{flushright}
KUCP0190\\
\end{flushright}
\vskip 4cm
\begin{center}
{\Large \bf
  Branes for Relativists  
}\\
\vskip 2cm
{  Jiro Soda\footnote{E-mail:jiro@phys.h.kyoto-u.ac.jp} \\
\vskip 0.5cm
  Department of Fundamental Science, Faculty of Integrated Human Studies, \\
  Kyoto University, Kyoto, 606--8501, Japan
}
\end{center}

\vskip 5cm
\begin{center}
{\large Abstract}
\end{center}
 Recently, branes in supergravity have become an indispensable tool even for 
traditional relativists. The purpose of this manuscript is to provide a pedagogical account of the brane technology so that the relativists can use branes
 in their study. The type IIA supergravity theory is mainly discussed and 
 other cases are briefly mentioned. The repulson singularity is also explained
 as an interesting application.

\newpage
\section{Introduction}

 Let me start to review the framework of the string theory briefly~\cite{pol,GSW,kiri}. 
The perturbative string theory can be formulated as the path-integral 
 of the Poyakov action over all the Euclidean world sheet geometry 
which has the non-trivial topology characterized by the Euler number
\begin{equation}
   \int DX Dh_{ab} \exp[ - {1\over 4\pi \alpha'} \int d^2\sigma 
    \sqrt{h} h^{ab} \partial_a X^\mu \partial_b X_\mu  ] \ ,
\end{equation}
where $X^\mu$ and $h_{ab}$ are the 10-dimensional coordinates specifying
 the string location and the induced metric on the world sheet of the
 string, respectively. 
 The complicated world sheet topology corresponds
 to the loop Feynmann diagram in the conventional field theory.
 In particular, the simplest one, i.e. the sphere, corresonds
 to the tree diagram. Hence the string clasical theory can be obtained by 
 path-integrating  the Poyakov action  on the sphere. 
 This includes the effects of all of the stringy excitations. 
 In the low energy limit, only the massless states can contribute 
 to the effective action. The stringy corrections can be incorporated 
 systematically like as 
\begin{equation}
 I = \int d^{10}x \sqrt{-g} e^{-2\phi} \left[
            \{ R + 4 (\partial \phi )^2 + \cdots \}
        + l_s^2 \{  R^{\mu\nu\lambda\rho} R_{\mu\nu\lambda\rho} + \cdots \} 
                                      \right]     \ ,
\end{equation}
where $l_s = \sqrt{\alpha'}$ is the string length scale and $g=e^{\phi}$ 
 controles the strength of the string coupling. 
Notice that the leading term
 is essentially that of the general relativity. From now on, we shall 
concentrate ourselves to the anlalysis of this part. 
 The validity of the analysis is garanteed if the string coupling constant 
 and the typical energy scale satisfy the inequality $g\ll 1 $ and 
$l_s E \ll 1 $.  Namely, we consider  the low energy classical solution 
of string theory. However, as the extreme branes are BPS objects, 
they are related to the non-perturbative objects in string theory, i.e.
 D-branes. This is the reason why the extreme branes are investigated 
 enthusiastically. 

 The organization of this review is as follows: In sec.2, the extreme 
Reissner-Nordstrom black hole solution in general relativity is derived
 in an appropriate manner for later purpose. In sec.3, general brane solutions
 are explained. In sec.4, we review the basic branes in type IIA theory. 
 Among them, Dp-branes coupled with R-R gauge fields are important 
 from the string theoretical point of view. 
 In sec.5, we consider the intersecting branes. In sec.6, the repulson 
singularity is explained. In the final section, we discuss the brane solutions
 in other theories.  In this review, we will not intend to cite the 
original papers
 because of the pedagogical nature of this note. Instead, we refer the readers
 to excellent review works~\cite{marolf,peet,argurio,stelle}.

\section{Extreme Reissner-Nordstrom (R-N) in General Relativity }

The charged black hole solution is familiar to relativists. Although its 
extreme limit has less importance in relativity, its stringy counterpart has 
taken an important role in the recent development of the nonperturbative 
aspects of string theory. Especially, the success of the statistical counting
 of the black hole entropy was impressing. The extreme R-N black hole solution
 would be a natural starting point for understanding the brane physics. 
The extreme R-N solution is obtained from the action: 
\begin{equation}
 S=  {1\over 16\pi G_4 } \int d^4 x \sqrt{-g} [R - {1\over 4} F_2^2 ] 
\end{equation}
where $g_{\mu\nu}$ is the 4-dimensional metric, 
$F_2 = dA$ is the electro-magnetic field strength. 
In this setting, it is possible to find a solution for which the test charged 
particle  does not feel any force. The ansatz is
\begin{equation}
 ds^2 = - H(r)^{-2} dt^2 + H(r)^2 ( dr^2 + r^2 d\Omega^2 )  \ \ ,
   \ \ A_t = H(r)^{-1} -1  \ ,
\end{equation}
where (0+1)-dimensional Poincare symmetry and the $SO(3)$-symmetry are
 imposed. The action for the test brane is defined as
\begin{equation}
 S_{particle}=   e \int A_\mu dx^\mu -e \int d\tau 
\end{equation}
and the
charged particle has the extreme mass, namely $e=m$ in our units.
The no force property can be verified by evaluating the above action
 as 
\begin{equation}
 S_{particle}=  e \int (H^{-1} -1 ) - e \int \sqrt{H^{-2} - H^2 v^2} dt 
  \sim e \int H^3 v^2     \ . 
\end{equation}
Here, the potential terms are canceled out. 
This type of the solution is sometimes represented by the following table:
\bigskip
\begin{center}
\begin{tabular}{|c|c|c|c|c|}
\hline
&$x^0$&$x^1$&$x^2$&$x^3$\\\hline
RN&$\bullet$&$-$&$-$&$-$\\\hline
\end{tabular}
\end{center}
\bigskip
 We may call this solution 0-brane, because it has no spatial extension.

Given the mtric form, the  curvature is easily calculated as
\begin{equation}
 R = -2 H^{-3} \nabla^2 H \ .
\end{equation}
Then, the vacuum Einstein equation reduces to the harmonic equation 
\begin{equation}
  T^\mu_\mu =0 \Rightarrow R=0 \Rightarrow \nabla^2 H =0  \ .
\end{equation}
It is easy to write down the general solution, i.e. multi-black hole solution.
 Here we consider the simplest solution 
\begin{equation}
    H = 1 + {Q \over r}   \ .
\end{equation}
In this case, the metric becomes
\begin{equation}
 ds^2 = - (1+{Q\over r})^{-2} dt^2 + (1+ {Q\over r})^2 (dr^2 + r^2 d\Omega^2 ) 
\end{equation}
or, by using the coordinate $R=r+Q$, 
\begin{equation}
 ds^2 = -(1-{Q\over R})^2 dt^2 + {dR^2 \over (1-{Q\over R})^2 } 
                              + R^2 d\Omega_2^2     \ .
\end{equation}
This is the well known form of the extreme R-N black hole solution.

We have discussed the electric charge only. 
Because the magnetic charge does not lead to the new metric solution
 due to the following simple relation
\begin{equation}
 F = dA \ , \ \     { }^* F = d\tilde{A}   \ .
\end{equation}
As we will see, however, the electric solution and the magnetic solution 
 are different in the case of general models.

\section{Branes:general}

Here, we would like to generalize the extreme Reissner-Nortstrom solution to 
 the higher dimensional gravity with the dilaton. One new thing comes here.
  The higher rank gauge fields are allowed to couple with the dilaton.

The action is
\begin{equation}
  S= {1\over 16\pi G_D } \int d^D x \sqrt{-g} \{ R - {1\over 2} \partial_\mu 
   \phi \partial^\mu \phi - {1\over 2} {1\over (p+2)! } e^{a\phi} F^2_{p+2} \}
\end{equation}
where $g_{\mu\nu}$ is the $D$-dimensional metric, $\phi$ is the dilaton, 
 and $F_{p+2} = dA_{p+1}$ is the (p+2)-form field strength. 
 The coupling constant $a$ is a key parameter in the brane physics. 
 It reflects the origin of the fields, namely NS-NS or R-R, and the rank
 of the tensor field. 
 In order to find the brane solutions, we need to impose the 
(p+1)-dimensional Poincare invariance and $SO(d)$-symmetry of the transverse
 space to the brane. Here, $d$ is the co-dimensions of the brane.
 Moreover, as we seek the extreme solution,  we need to consider the
 no force condition. The action of the test brane is given by
\begin{equation}
  S_{brane} =  \mu_p \int A_{p+1} 
     - \tau_p \int d^{p+1} x f(\phi ) \sqrt{- \det G_{ab}} \ ,
\end{equation}
where $\mu_p$ and $\tau_p$ are the  charge and the tension of the brane
 and $G_{ab}$ is the induced metric on the brane.
 The extreme solution is defined by $\mu_p = \tau_p $. 
Here, $f(\phi) $ takes the  form 
\begin{equation}
    f(\phi ) = e^{{p-3 \over 4} \phi } 
\end{equation}
for the R-R forms and
\begin{equation}
     f(\phi ) = e^{{p+1 \over 4} \phi }
\end{equation}
for NS-NS forms in the type II supergravity theories. 
 In the case of the M-theory, as there exists no dilaton, 
 this part of the action is trivial. 
The ansatz we imposed is
\begin{eqnarray}
 ds^2 &=& H^{-2{d-2 \over \Delta}} \{-dt^2 + dy_1^2 +\cdots + dy_p^2 \}  
      + H^{2{p+1 \over \Delta }} \{ dr^2 + r^2 d\Omega_{d-1}^2 \}  \ , \\
    e^{\phi} &=& H^{a(D-2) \over \Delta} \ , \\
    A_{ty_{1} \cdots y_{p}}  &=& \sqrt{2(D-2) \over \Delta} 
                                              ( H^{-1} -1 ) \ , \\
    \Delta &=& (p+1) (d-2) + {1\over 2} a^2 (D-2)  \ ,
\end{eqnarray}
where $H$ depends only on $r$ and $D=p+d+1$. 
It is easy to verify the no force property of the brane for each model 
 separately. We leave it as an exercise for the reader. 
 After the straightforward but tedious calculation, we have the harmonic
 equation 
\begin{equation}
    \nabla^2 H = 0  \ .
\end{equation}
Its solution is given by 
\begin{equation}
      H = 1 + {h^{d-2} \over r^{d-2} }
\end{equation}
where the integration constant $h$ is related to the charge of the brane.
So far, we have discussed the electric solution. The following relation
 suggests the property of the magnetic solution,
\begin{equation}
      F_{p+2} = dA_{p+1} \ ,  \ \ \ \ { }^{*} F_{D-p-2} = dA_{D-p-3} 
\end{equation}
Namely, the magneteic solution is nothing but $D-p-4$ brane solution.

\section{Basic branes in type IIA SUGRA }

The low energy effective action of the type IIA superstring theory is 
the type IIA supergravity theory. The bosonic part of the 
action in the string frame is given by
\begin{equation}
     S= {1\over 16\pi G_{10} } \int d^{10} x \sqrt{-g_{s}} \{e^{-2\phi} \left[ R 
     +4\partial_\mu \phi \partial^\mu \phi - {1\over 12}  H^2_{3}  \right]
      - {1\over 4}  F^2_{2}  - {1\over 48}  F^2_{4}  \} 
\end{equation}
The transformation rule 
\begin{equation}
      g^{s}_{\mu\nu} = e^{\phi \over 2} g^{E}_{\mu\nu} 
\end{equation}
gives the action in the Einstein frame
\begin{equation}
      S= {1\over 16\pi G_{10} } \int d^{10} x 
       \sqrt{-g_{E}} \{  R - {1\over 2} \partial_\mu \phi \partial^\mu \phi 
      - {1\over 12} e^{-\phi}  H^2_{3}  
      - {1\over 4} e^{{3\over 2}\phi} F^2_{2}  
      - {1\over 48} e^{{1\over 2}\phi} F^2_{4}  \}  \ .
\end{equation}
Now, we can use the results of the previous section in order to
 obtain the brane solutions. In this section, we consider the single
 brane, and the intersecting branes are the subjects of the next section.

\subsection{NS-NS sector}

 The gauge field in the NS-NS sector is
\begin{equation}
     H_3 = dB_2  \ .
\end{equation}
The antisymmetric tensor field $B_2 $ naturally couples 
with the fundamental string.
From the action one can read off $a=-1$ and then
obtain $\Delta = 2\cdot 6 + 1\cdot 8 /2 =16 $.
Notice that $D=10$ in superstring theories. 
The electric brane is the fundamental 1-brane:
\bigskip
\begin{center}
\begin{tabular}{|c|c|c|c|c|c|c|c|c|c|c|}
\hline
&$x^0$&$x^1$&$x^2$&$x^3$&$x^4$&$x^5$&$x^6$&$x^7$&$x^8$&$x^9$\\\hline
F1&$\bullet$&$\bullet$&$-$&$-$&$-$&$-$&$-$
&$-$&$-$&$-$\\
\hline
\end{tabular}
\end{center}
\bigskip
The solution is given by
\begin{eqnarray}
  ds^2_{F1} &=& H^{-{3\over 4}}(-dt^2 + dy_1^2 ) + H^{{1\over 4}} 
      (dx_1^2 + \cdots + dx_8^2 )  \ , \\ 
   e^{\phi} &=& H^{-{1\over 2}} \ , \\
   B_{ty_{1}}  &=& H^{-1} -1 \ , \\
    H &=& 1+ {h^6 \over r^6}     \ .  
\end{eqnarray}
It is easy to obtain the metric in the string frame:
\begin{equation}
   d\hat{s}^2_{F1}  = H^{-1} (-dt^2 + dy_1^2 ) +  dx_1^2 + \cdots + dx_8^2   \ .
\end{equation}
The magnetic brane is NS 5-brane:
\bigskip
\begin{center}
\begin{tabular}{|c|c|c|c|c|c|c|c|c|c|c|}
\hline
&$x^0$&$x^1$&$x^2$&$x^3$&$x^4$&$x^5$&$x^6$&$x^7$&$x^8$&$x^9$\\\hline
NS5&$\bullet$&$\bullet$&$\bullet$&$\bullet$&$\bullet$&$\bullet$&
$-$&$-$&$-$&$-$
\\
\hline
\end{tabular}
\end{center}
\bigskip
The metric is given by
\begin{eqnarray}
    ds^2_{NS5} &=& H^{-{1\over 4}}(-dt^2 + dy_1^2 +\cdots +dy_5^2 ) 
   + H^{{3\over 4}}   (dx_1^2 + \cdots + dx_4^2 )  \ ,\\ 
    e^{\phi} &=& H^{{1\over 2}}  \ , \\
    H_{\theta_1, \theta_2 , \theta_3} &=& Q \omega_3  \ , \\
    H &=& 1 + {h^2 \over r^2}     \ .
\end{eqnarray}
Here, $Q$ is related to the other  constant $h$. 
In the string frame, the  metric becomes
\begin{equation}
      d\hat{s}^2_{NS5}  = -dt^2 + dy_1^2 +\cdots +dy_5^2 
                         +  H (  dx_1^2 + \cdots + dx_4^2 ) \ .
\end{equation}

\subsection{R-R sector}

The gauge fields in the R-R sector are
\begin{equation}
      F_2 = dC_1 
\end{equation}
and
\begin{equation}
      F_4 = dC_3  \ .
\end{equation}
From the action, we can find $a_p = (3-p)/2 $ and then 
$\Delta = (p+1)(7-p)+ (3-p)^2 =16$.
 Then, the solution becomes
\begin{eqnarray}
    ds^2_{Dp} &=& H^{-{7-p \over 8}}(-dt^2 + dy_1^2 +\cdots +dy_p^2 ) 
   + H^{{p+1 \over 8}}   (dx_1^2 + \cdots + dx_{9-p}^2 )  \ , \\ 
    e^{\phi} &=& H^{{3-p \over 4 }}  \ , \\
    C_{t y_{1} \cdots y_{p} } &=& H^{-1} -1 \ , \\ 
    H &=& 1 + {h^{7-p} \over r^{7-p} }     \ .
\end{eqnarray}
Also, the metric in the string frame is
\begin{equation}
      d\hat{s}^2_{Dp}  = H^{-{1\over 2}} ( -dt^2 + dy_1^2 +\cdots +dy_p^2 ) 
                   +  H^{1\over 2} (  dx_1^2 + \cdots + dx_{9-p}^2 ) \ .
\end{equation}

\section{Harmonic Superposition}

Interestingly, based on the previous results, one can derive various
 solutions without calculations. The superposition rule is the only
 necessary ingredient for this purpose. 
 For example, in the case of the D-brane, we obtain the following 
 superposition rules
\begin{equation}
   Dp \cap Dp = p-2 \ , \ \ D(p-2) \cap Dp = p-3 \ , 
  \ \   D(p-4) \cap Dp = p-4   \ ,
\end{equation}
where the right hand side represents the number of the common spatial
 dimensions. 
These rules can be deduced from the consideration of the supersymmetry.
 The extreme brane preserves one half of the space-time supersymmetry
 of the theory. So the problem is how to combine the branes so as not to
 break all of the supersymmetry of the theory. In the case of the Dp-branes,
 we obtained the above rules. Of course, these rules can be obtained by 
 the elementary analysis of the equations of motion. 

As an illustration, we consider the D6-D2 system:
\bigskip
\begin{center}
\begin{tabular}{|c|c|c|c|c|c|c|c|c|c|c|}
\hline
&$x^0$&$x^1$&$x^2$&$x^3$&$x^4$&$x^5$&$x^6$&$x^7$&$x^8$&$x^9$\\\hline
D6&$\bullet$&$\bullet$&$\bullet$&$\bullet$&$\bullet$&$\bullet$&
$\bullet$&$-$&$-$&$-$\\\hline
D2&$\bullet$&$\bullet$&$\bullet$&$|-|$&$|-|$&$|-|$&
$|-|$&$-$&$-$&$-$
\\
\hline
\end{tabular}
\end{center}
\bigskip
As $D6\cap D2 =2$, the $1, 2$-directions are chosen as the 
 common directions.
The solution is obtained by the simple superposition of both solutions
\begin{equation}
      d\hat{s}^2_{Dp}  = H_2^{-{1\over 2}} H_6^{-{1\over 2}} 
                   ( -dt^2 + dy_1^2  +dy_2^2 ) 
                   + H_2^{{1\over 2}} H_6^{-{1\over 2}} 
                   ( dy_3^2 + dy_4^2  +dy_5^2+dy_6^2 ) 
       +  H_2^{{1\over 2}} H_6^{{1\over 2}}  (  dr^2  + r^2 d\Omega_2^2 ) 
\end{equation}
where
\begin{equation}
     H_2 = 1 + {r_2 \over r }  \ ,  \ \ H_6 = 1+ {r_6 \over r}   \ .
\end{equation}
Notice that the power of the harmonics is determined by the dimensions of
 the transverse spaces to all of the branes. 
This is the so-called smearing effects.

\section{Repulson Singularity}

It would be interesting to study the wrapped brane $D6-D2^{*}$, 
because it has a naked time-like singularity and its resolution 
 in the string theoretical context is known.  Notice that the 
 four-dimensional manifold K3 is supersymmetric itself, hence
 the wrapped D6 brane solution  on K3 is also supersymmetric. 
 The resulting solution induces negative charged D2 branes.
 Then, this system is often called $D6-D2^{*}$. 
 The metric is given by 
\begin{equation}
     d\hat{s}^2_{Dp}  = H_2^{-{1\over 2}} H_6^{-{1\over 2}} 
                   ( -dt^2 + dy_1^2  +dy_2^2 ) 
       +  H_2^{{1\over 2}} H_6^{{1\over 2}}  (  dr^2  + r^2 d\Omega_2^2 ) 
       + V^{{1\over 2}} H_2^{{1\over 2}} H_6^{-{1\over 2}} ds_{K3}^2  \ ,
\end{equation}
where
\begin{equation}
     H_2 = 1 - {r_2 \over r }  \ , \ \  H_6 = 1+ {r_6 \over r}  \ .
\end{equation}
Here, $ds_{K3}^2$ is the metric of a K3 surface of unit volume. 
 V is the volume of the K3 as measured at infinity, but the 
 supergravity solution adjusts itself such that 
$V(r) = V H_2 H_6^{-1}$ is the measured volume of K3 at radius $r$.
 Other fields such as R-R forms can be obtained easily. 
Considering the connection with the string theory, we obtain
\begin{equation}
     r_2 = {(2\pi)^4 gN \alpha'^{5\over 2} \over 2 V}  \ , 
    \ \  r_6 = {8 N \alpha'^{1\over 2} \over 2} \ .
\end{equation}
There is a time-like naked singularity at $r=r_2$, known as
 ``repulson" (see Fig.1) .
\begin{figure}[t]
  \epsfysize=50mm
   \epsfbox{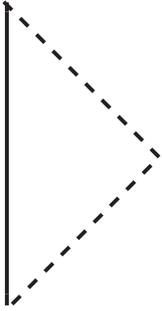}
  \caption{Penrose diagram for $D6-D2^{*}$ solution. The bold line represents
 the time-like naked singularity. }
\label{penrose-1}
\end{figure}
The curvature diverges there which is related to
 the fact that the volume of the K3 goes to zero there.

Let us see the repulsive nature of this solution. 
The Newtonian potential is given by
\begin{equation}
     \Phi = {1\over 2}\left[ - g_{00} -1 \right]
          = {1\over 2}\left[ (1-{r_2 \over r})^{-{1\over 2}} 
            (1+ {r_6 \over r})^{-{1\over 2}} -1 \right] 
\end{equation}
Let $r\rightarrow \infty $, then
\begin{equation}
     \Phi \approx {r_2 - r_6 \over 4r} 
          \approx - {gN\alpha'^{1\over 2} \over 8r} 
                    \left[1-{V^{*} \over V} \right]
\end{equation}
thus, at the infinity, the gravity force is attractive as is expected. 
Here we assumed  
\begin{equation}
     V^{*} = (2\pi )^4 \alpha'^2 < V \ .
\end{equation}
As you can see, however, the potential becomes repulsive near the naked
 singularity. The resolution of this singularity will be  discussed
 in the review talk by Yamaguchi.

\section{Others}

The method explained in this review can be easily extended to
 the other theories.

In the case of 11 dimensional supergravity, the bosonic part of the
 action is given by 

\begin{equation}
      S= {1\over 16\pi G_{10} } \int d^{11} x 
       \sqrt{-g} \{  R   
      - {1\over 48}  H^2_{4}  \}  \ ,
\end{equation}
where no dilaton exists. The possible solutions are
\begin{equation}
     W \ , KK  \ , M2 \ , M5  \ .
\end{equation}
The Kalza-Klein wave solution is
\begin{eqnarray}
ds^2_{W} &=&  -dt^2 + dy^2 + K (dt-dy)^2 
   + dx_1^2 + \cdots + dx_9^2  \ , \nonumber \\
   & & K = {Q \over r^7}    \ . 
\end{eqnarray}
The Kalza-Klein monopole solution is
\begin{eqnarray}
ds^2_{KK} &=&  -dt^2 + dy_1^2 +\cdots +dy_6^2  
   + H^{-1} (dz + A_i dx^i )^2 + H dx^i dx_i  \ , i= 1...3  \ , \nonumber \\
   & & H = 1 + {Q\over r} \ , \ \ 
   \partial_i A_j - \partial_j A_i = - \epsilon_{ijk} \partial_k H \ .
\end{eqnarray}
Both of the above solutions have off-diagonal components, then
 they do not belong to the category discussed in the previous sections.
However, they can be derived by reversing the Kalza-Klein procedure to
 derive 11-dimensional metric from the 10-dimensional brane solution. 

The electric 2-brane for the 3-form field can be obtained 
 by setting $a=0$ in the general brane solution. The result is 
\begin{eqnarray}
ds^2_{M2} &=& H^{-{2\over 3}}(-dt^2 + dy_1^2 +dy_2^2 ) 
   + H^{{1\over 3}}   (dx_1^2 + \cdots + dx_8^2 ) \ , \nonumber \\
     & &  H = 1 + {Q\over r^6} \ .
\end{eqnarray}
 Similarly, the magnetic 5-brane is deduced as
\begin{eqnarray}
ds^2_{M5} &=& H^{-{1\over 3}}(-dt^2 + dy_1^2 +\cdots +dy_5^2 ) 
   + H^{{2\over 3}}   (dx_1^2 + \cdots + dx_5^2 )  \ , \nonumber \\
    & & H = 1 + {Q \over r^3}  \ .
\end{eqnarray}

In the case of the type IIB supergravity, the bosonic part of 
 action in the string frame is
\begin{equation}
     S= {1\over 16\pi G_{10} } \int d^{10} x \sqrt{-g_{s}} \{e^{-2\phi} \left[ R 
     +4\partial_\mu \phi \partial^\mu \phi - {1\over 12}  H^2_{3}  \right]
       - {1\over 2}  F^2_{0} - {1\over 12}  F^2_{3}  
                          - {1\over 240}  F^2_{5}  \} 
\end{equation}
and the action in the Einstein frame is 
\begin{equation}
      S= {1\over 16\pi G_{10} } \int d^{10} x 
       \sqrt{-g_{E}} \{  R - {1\over 2} \partial_\mu \phi \partial^\mu \phi 
      - {1\over 12} e^{-\phi}  H^2_{3}  
      - {1\over 2} e^{2\phi}  F^2_{0} 
      - {1\over 12} e^{\phi} F^2_{3}  
      - {1\over 240}  F^2_{5}  \}    \ .
\end{equation}
In this case, the possible branes are
\begin{equation}
    W \ , KK  \ , F1 \ , NS5 \ , D(-1) \ , D1 \ , D3 \ , D5 \ .
\end{equation}
Compare with the case of the type IIA supergravity
\begin{equation}
     W \ , KK  \ , F1 \ , NS5 \ , D0 \ , D2 \ , D4 \ , D6  \ .
\end{equation}
As the formula for the metric in type IIB theory is almost the same as  
 those in type IIA theory, we do not display them here.

\end{document}